\documentclass[final,twocolumn,rmp,floats,amsfonts,amssymb,amsmath,superscriptaddress,floatfix,a4paper]{revtex4}
\usepackage{graphicx}
\usepackage{hyperref}
\usepackage{subfigure}
\usepackage{url}

\begin{document}
\title{Experimental studies of liquid-liquid dispersion in a turbulent shear flow.}
\author{Florent Ravelet}
\affiliation{Laboratory for Aero and Hydrodynamics, Leeghwaterstraat 21, 2628 CA Delft, The Netherlands.}
\email{florent.ravelet@ensta.org}
\author{Ren\'e Delfos Jerry Westerweel}
\affiliation{Laboratory for Aero and Hydrodynamics, Leeghwaterstraat 21, 2628 CA Delft, The Netherlands.}
\author{Jerry Westerweel}
\affiliation{Laboratory for Aero and Hydrodynamics, Leeghwaterstraat 21, 2628 CA Delft, The Netherlands.}
\maketitle

\section{abstract}
We study liquid-liquid dispersions in a turbulent Taylor--Couette flow, produced between two counterrotating coaxial cylinders. In pure Water and in counterrotation, Reynolds numbers up to $1.4 \times 10^5$ are reached. The liquids we use are a low-viscous Oil and pure Water or a Sodium Iodide solution with a refractive index matched to that of Oil, in order to get transparent dispersions. We first characterize the single-phase flow, in terms of threshold for transition to turbulence, scaling of the torque and measurements of the mean flow and of the Reynolds stress by stereoscopic PIV. We then study the increase of the dissipation in the two-phase flows and find that the torque per unit mass can be twice the torque for a single-phase flow. Long-time behaviours are also reported. 

\section{Introduction}

Liquid-liquid dispersions are encountered for instance in extraction or chemical engineering when contact between two liquid phases is needed. Without surfactants, when two immiscible fluids are mechanically agitated, a dispersed state resulting from a dynamical equilibrium between break-up and coalescence of drops can be reached. Their modelling in turbulent flows is still limited (Portela and Oliemans, 2006).

Some open questions we address are first the potential increase of turbulent dissipation due to the dispersion, the modification
of the turbulent transport inside the flow, and the droplet size distribution, governed by the breakup and coalescence rates. Finally, some 
concentrated liquid-liquid flows are also known to experience hysteretic phase inversion (Piela et al., 2006). 

To answer these questions, we have chosen a model shear-flow: a Taylor--Couette flow where the fluids are stirred in the gap between two counterrotating coaxial cylinders. We use low-viscous fluids and first check that the single-phase flows are strongly turbulent. We measure the wall shear stress, and use Refractive-Index matched fluids in order to measure the phases repartition.

\section{Experimental setup}
\begin{figure}[htb]
\centering
\includegraphics[width=.45\textwidth]{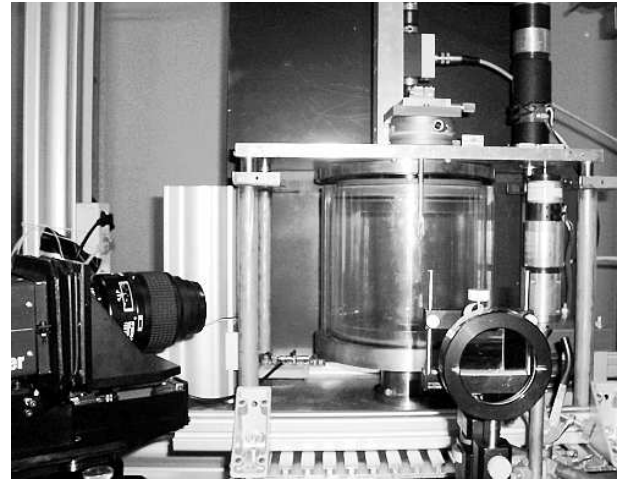}
\caption{Experimental setup, with the rotating torquemeter (upper part of picture), the calibration grid displacement device (on top of the upper plate), one of the two cameras (left side) and the light sheet arrangement (right side). The second camera is further to the right.}
\label{fig:manip}
\end{figure}

The flow is produced between two coaxial cylinders. The inner one is of radius $r_i=110 \pm 0.05$mm, and the outer one of radius $r_o=120 \pm 0.05$mm, which gives a gap ratio $\eta=r_i/r_o=0.917$. The length of the cylinders is $L=220$mm, i.e. the axial aspect ratio is $L/(r_o-r_i)=22$. The cylinders axis is vertical in order to keep axisymmetric conditions. The system is closed at both ends, with top and bottom lids rotating with the outer cylinder. There is a gap of $1.5$mm between these lids and the flat ends of the inner cylinder. The cylinders are driven by two independent Brushless DC motors (Maxon, $250$W), at a speed up to $10$Hz. The torque $T$ exerted on the inner cylinder is measured with a HBM rotating torquemeter (T20WN, $2$N.m). A photograph of the setup is presented in figure~\ref{fig:manip}. The optical measurements have been done with an outer glass cylinder, of radius $121 \pm 0.25$mm ($\eta=0.909$).

For a newtonian fluid of kinematic viscosity $\nu$, we use the set of parameters defined by Dubrulle et al. (2005): a mean Reynolds number $Re=2/(1-\eta) \; |(\eta r_o \omega_o d /\nu)- (r_i \omega_i d /\nu)|$ based on the shear and on the gap $d$; and a \lq\lq Rotation number\rq\rq~$Ro$ which is zero in case of perfect counterrotation ($r_i \omega_i= - r_o \omega_o$). At $10$Hz in counterrotation, the typical shear rate is around $1400$s$^{-1}$ and $Re \simeq 1.4 \times 10^5$ for pure Water. 

\begin{table}[htb]
\caption{Physical properties of the fluids at $20^o$C: density $\rho$ (kg.m$^{-3}$), dynamic viscosity $\mu$ (Pa.s) and kinematic viscosity $\nu$ (m$^2$.s$^{-1}$).}
\vspace{3mm}
\centering
\begin{tabular}{lccc}
\hline\hline
 & $\rho$ & $\mu$ & $\nu$ \\
\hline
Water & $1000$ & $1.0 \times 10^{-3}$ & $1.0 \times 10^{-6}$\\
NaI solution & $1500$ & $2.0 \times 10^{-3}$ & $1.3 \times 10^{-6}$\\
Oil & $800$ & $3.0 \times 10^{-3}$ & $3.8 \times 10^{-6}$\\
\hline\hline
\end{tabular}
\label{tab:fluids}
\vspace{6mm}
\end{table}

The physical properties of the fluids we use are given in table~\ref{tab:fluids}. The Oil is the Shell Macron EDM110; it is a low-viscosity paraffinic hydrocarbons mix. Its refractive index measured with an Abbe refractometer at $20^o$C is $1.4445$. The interfacial tension between Oil and Water is $\sigma=0.045$N.m$^{-1}$. To have a refractive index matched dispersion (Budwig, 1994), in addition to pure Water, we also use a Sodium Iodide solution as acqueous phase (Narrow et al., 2000). The concentration of NaI is $510$g.L$^{-1}$, and traces of Na$_2$S$_2$O$_3$ are added to avoid yellowish coloration. The viscosity ratio is then close to unity and the density ratio is close to $2$. In the following, unless specified, dimensionless quantities are deduced from $r_i$ as unit length, $1/\omega_i$ as unit time, and $\rho r_i^3$ as unit mass.


\section{Results}
\subsection{Single-phase flow}
\subsubsection{Torque scaling}
\begin{figure}[htb]
\centering
\includegraphics[width=75mm]{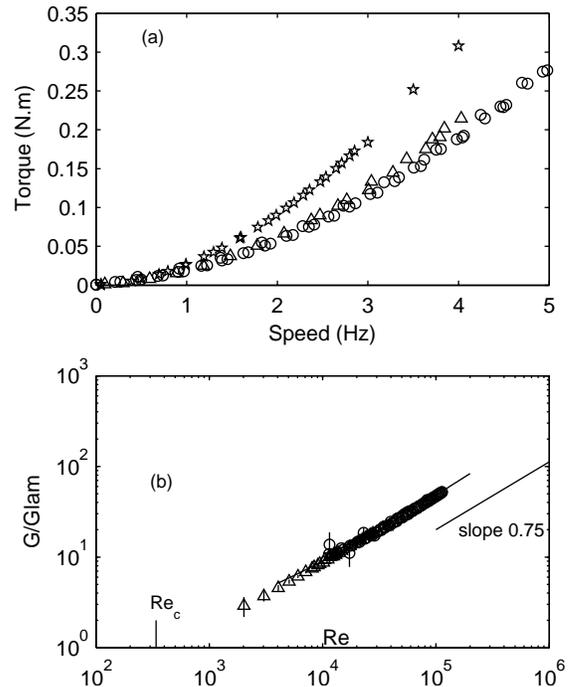}
\caption{(a): Torque {\em vs.} rotation frequency in exact counterrotation for Oil ($\triangle$), Water ($\circ$) and NaI ($\star$). (b): Dimensionless torque $G$ normalised by laminar torque $G_{lam}$ {\em vs.} $Re$. Solid line is the fit proposed by Dubrulle and Hersant (2002). $Re_c$ is the expected value for the first instability threshold, according to the theory of Esser and Grossmann (1996).}
\label{fig:pure}
\end{figure}

We first study the turbulent Taylor-Couette flow in exact counterrotation, and characterize its turbulent state. Though the Taylor-Couette flow has been widely studied, few experimental results and theories are available for $Ro=0$ (Dubrulle et al., 2005). We report torque measurements on pure fluids in figure.~\ref{fig:pure}. They have been made in a first version of the experiment, with a free surface and a space of $10$mm between the cylinders bottom ends. The torque due to the bottom part has been removed by studying different filling levels. The range of Reynolds numbers we cover is $2.0 \times 10^3 \lesssim Re \lesssim 1.2 \times 10^5$. In figure~\ref{fig:pure}a, we plot the dimensional torque \emph{vs.} speed for the three pure fluids. The torques behave non-linearly with speed, as expected for turbulent regimes, and the Oil and Water are very close, the higher viscosity of Oil balancing its lower density. We plot in figure~\ref{fig:pure}b the dimensionless torque $G=T/(\rho \nu^2 L)$ normalised by the laminar dimensionless torque $G_{lam}=2\pi \eta / (1-\eta)^2 Re$ \emph{vs.} $Re$. The experimental data fall on the same curve, which is consistent with the formula proposed by  Dubrulle and Hersant (2002) of the form $G=a\,Re^2/(ln(b\,Re^2))^{(3/2)}$. Here, a non-linear fit gives $a=16.5$ and $b=7\times 10^{-5}$. At the end of the range, this is close to a power-law behaviour in $G \propto Re^{1.75}$. This scaling indicates that in counterrotation, a turbulent regime is easily reached (let us remind that the scaling expected in the Taylor-vortices regime is $G \propto Re^{1.5}$).

Moreover, Esser and Grossmann (1996) proposed a formula for the first instability threshold. Here it gives $Re_c(\eta,Ro)=338$. At first glance, this value seems to be consistent with our results (see figure~\ref{fig:pure}b). In forthcoming experiments, we will use glycerol to lower the Reynolds number in order to check this prediction.

\subsubsection{Stereoscopic PIV measurements}
We measure the three components of the velocity in one plane by stereoscopic PIV. The plane, illuminated by a double-pulsed Nd:YAg laser is normal to the mean flow, i.e. vertical (see figure~\ref{fig:manip}): the in-plane components are the radial ($u$) and axial ($v$) velocities, while the out-of-plane component is the azimuthal component ($w$). It is imaged using two double-frame PCO-cameras, mounted on Scheimpflug adaptors, located at each side of the light sheet, with an angle of $60^o$ in the air. The flow is seeded with tracer particles (8 micrometer Sphericel). The field of view measures $11$mm $\times$ $22$mm, corresponding to a resolution of $300 \times 1024$ pixels. 

Special care have been taken concerning the calibration procedure on which the third azimuthal component rely. We use a very thin transparent sheet with crosses printed on it, with a weight at its bottom, and attached to a rotating and translating micro-traverse (see figure~\ref{fig:manip}). It is first put into the light sheet and traversed perpendicularly to it. Typically five pictures are taken with steps of $0.5$mm. The raw PIV-images are processed with PivWare (Westerweel, 1993), with a last interrogation area of $32 \times 32$ pixels with $50\%$ overlap, and normalised median filtering as post-processing. Then the vector mapping and the third component reconstruction are made with Matlab. The mapping function is a third-order polynomial, and interpolations are bilinear.

\begin{figure}[htb]
\centering
\includegraphics[width=75mm]{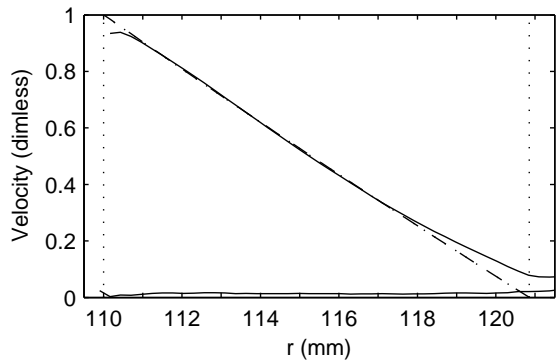}
\caption{Velocity profiles measured by stereoscopic PIV. Outer cylinder is at rest, $Re \simeq 90$. Solid line: measured mean azimuthal velocity. Dotted line: theoretical profile. Dashed line: fit of the form $w=ar+b/r$. The radial component $u$ which should be zero is also shown.}
\label{fig:pivgly}
\end{figure}

In order to check the quality of the technique, we measured the basic laminar flow with the outer cylinder at rest, for $\{ Ro=(1-\eta) \; ; \; Re \simeq 90 \}$. The measurements are done in $86 \%$ glycerol-water mixture. We plot in figure~\ref{fig:pivgly} the velocity radial profiles, averaged over $100$ recordings. The flow is stationary and laminar. It is in good agreement with the theoretical profile: $w(r)=(\eta^2/(\eta^2-1))\omega_i \; r \; + \; (1/(1-\eta^2))\omega_i r_i^2 \; 1/r$, except near the inner and outer walls, where some reflections disturb the measurement. The error in the in-plane velocity is roughly constant, around $0.015$. 

\begin{figure}[htb]
\centering
\includegraphics[width=75mm]{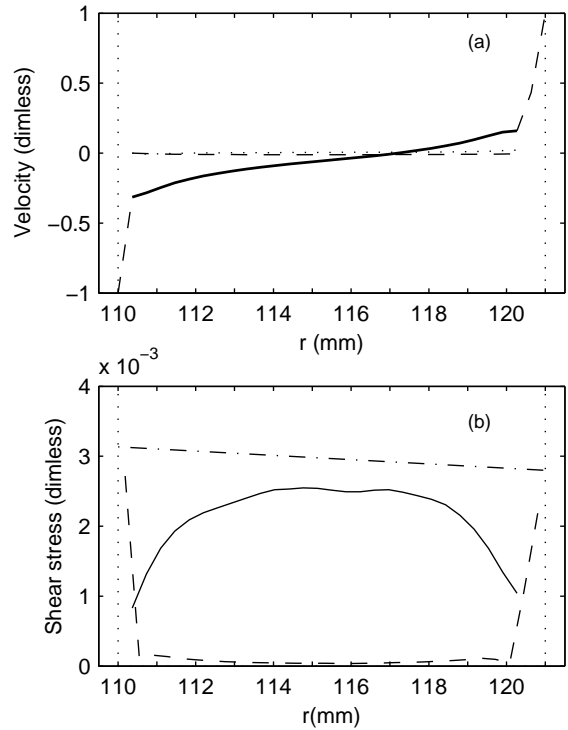}
\caption{Velocity measured by stereoscopic PIV for the counterrotating case, at $Re \simeq 1.4 \times 10^4$. (a): Dimensionless mean velocity profiles $u$ (dotted line), $v$ (dashed line), and $w$ (solid line). The azimuthal profile is not measured at the walls where its magnitude is $1$. (b): Dimensionless turbulent shear-stress $\langle u'w' \rangle$ (solid line) and dimensionless viscous shear-stress computed from the mean velocity profile. The dash-dotted line connects the two known values of the shear-stress at the walls, deduced from torque measurements.}
\label{fig:pivcontra}
\end{figure}

We then measured the counterrotating flow, at $Re=1.4 \times 10^4$. The results are presented in figure~\ref{fig:pivcontra}. The measurements are triggered on the outer cylinder position, and are averaged over $500$ images. In the counterrotating case, for this large gap ratio and at this value of the Reynolds number, the instantaneous velocity field is really desorganised and does not contain obvious structures like Taylor-vortices, in contrast with other situations (Wang et al., 2005). No peaks are present in the time spectra, and there is no axial-dependency of the time-averaged velocity field. We thus average in the axial direction the different radial profiles. The azimuthal velocity $w$ in the bulk is low, i.e. its magnitude is below $0.1$ between $112 \lesssim r \lesssim 120$ that is $75\%$ of the gap width. The two other components are zero within $0.002$.

The fluctuations of the velocity are not isotropic. They are of the order of $0.05$ for the radial and axial components, and of the order of $0.1$ for the azimuthal components. We can also compute some terms in the angular momentum balance equation (Mari\'e and Daviaud, 2004). According to the torque measurements described in the previous paragraph, the shear-stress at the inner wall is $\tau_i=T/(2\pi L r_i^2) \simeq 1.49$Pa, i.e. $\tau_i / (\rho r_i^2 omega_i^2)=3.1 \times 10^{-3}$ in dimensionless form. The dimensionless shear-stress at the outer wall is $2.8 \times 10^{-3}$. We report these two values in figure~\ref{fig:pivcontra}b, together with the viscous shear-stress $\mu r (\partial \omega / \partial r)$ (dashed line) and the turbulent shear stress $\rho \langle u'w' \rangle$ (solid line). The viscous shear-stress is negligible in the bulk, and is dominant at the walls, where it should ensure all the transport of angular momentum. Though the boundary layer is very badly resolved, the rough estimate of the viscous stress at the walls ---corresponding to the tails of the dashed line--- is consistent with the expected values. The measured turbulent shear-stress is roughly constant in the bulk, and has the good order of magnitude. The correlation coefficient between $u$ and $w$, defined as $\langle u'w' \rangle / \sqrt{\langle u'u' \rangle \langle w'w' \rangle}$ is $0.4$. The turbulent shear-stress represents $85\%$ of the expected shear-stress. We conclude that this flow is dominated by turbulent transport, which is also consistent with the high scaling of the torque with $Re$ (Remind that $G \propto Re^{1.75}$).

\subsection{Two-phase flow}
\subsubsection{Qualitative pictures of the dispersion process}
We now use two immiscible fluids in the Taylor-Couette facility.

\begin{figure}[htb]
\centering
\includegraphics[width=75mm]{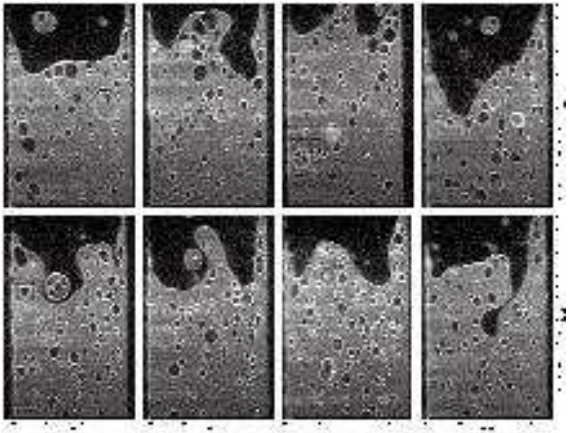}
\caption{LIF pictures of the dispersion in a counterrotating flow at $2$Hz. Dark region corresponds to the Oil, and bright regions to the Sodium Iodide solution. The outer cylinder is on the right side of each picture.}
\label{fig:lif}
\end{figure}

To look at the phases repartition, we perform visualisations with a Laser Induced Fluorescence technique. We add a small amount of Rhodamine B in the Refractive-index matched Sodium Iodide solution. The Rhodamine is sensitive to the green light of the Nd:YAg Laser ($532$nm), and emits a yellow fluorescent light. We use Melles-Griot low-pass filters, with a cutoff wavelength of $550$nm to only record the fluorescent light. Examples of visualisations are given in figure~\ref{fig:lif}.

Starting from fluids at rest, the system is organised into two layers. When the cylinders are rotating, there is a competition between centrifugal forces and gravity. A Froude number measuring the ratio of the centrifugal forces to buoyancy can be estimated as $Fr = (\frac{\rho u^2}{\Delta \rho g r_i})^{1/2} \simeq 0.9f$. We indeed observe that for $f\lesssim 1 $Hz the fluids stay separated, but that the interface is tilted, the heavier fluid being pushed outwards. The interface is very weavy and non stationary. The ratio of turbulent shear stress to buoyancy can be estimated as $Fr = (\frac{\rho u_{*}^2}{\Delta \rho g d})^{1/2} \simeq 0.23f$, with $u_{*}=0.075$ being the typical rms velocity measured by PIV. A Weber number comparing the effects of turbulent stresses to interfacial tension can be defined as 
$We=\frac{\rho u_{*}^2 d}{\sigma} \simeq 0.6f^2$. At a speed of $1.5$Hz, some Oil droplets start to be entrained into the acqueous phase. Very few drops of acqueous phase into Oil are present. Increasing further the speed, more and more droplets of Oil are entraped into NaI, and there start to be acqueous drops into the Oil, mainly visible as films, i.e. drops into drops, as can be seen in figure~\ref{fig:lif}. Then, there seems to be three regions in the fluids, from top to bottom: an Oil-continuous region with NaI drops, a foam of possibly multiple droplets, and a Nai-continuous region with a lot of Oil droplets. Finally, for $f \gtrsim 3.5$Hz, the fluids are fully dispersed.

\subsubsection{Evolution of the torque for dispersed phases}

One of the open question we adress is whether the turbulent dissipation is changed with respect to a monophasic case, and in that case, how much is that change. There are models for the effective viscosity of emulsions (Pal, 2001), but what does happen in a turbulent dispersion where molecular viscosity is usually negligible ?

\begin{figure}
\centering
\includegraphics[clip,width=75mm]{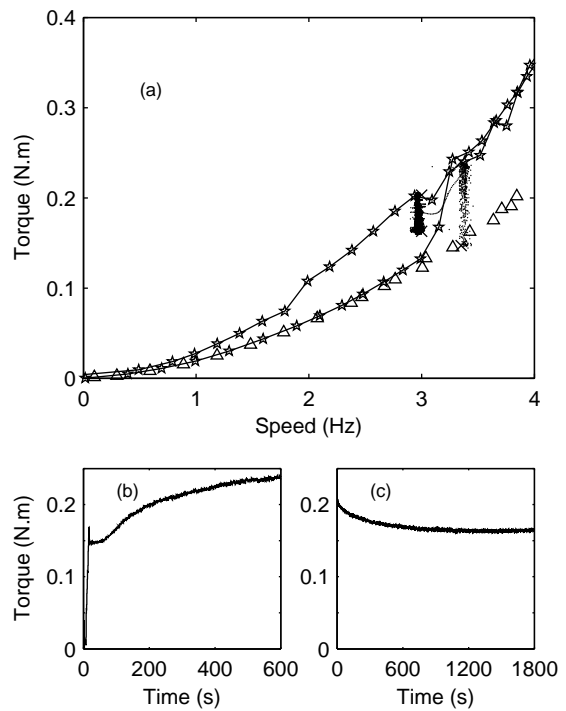}
\caption{(a): Torque \emph{vs.} rotation frequency for pure Oil ($\triangle$) and $33\%$ Water in Oil ($\star$). Speed is increased by steps of $0.2$Hz every $60$s, up to $4$Hz and then decreased back to $0$. The points clouds are parametric plots from experiments (b) and (c). (b): From a two-layer system at rest, the speed is set to $3.5$Hz. (c): From this speed, after 20 minutes, the speed is decreased to $3$Hz.}
\label{fig:dispersion}
\end{figure}

We plot in figure~\ref{fig:dispersion}a the torque \emph{vs.} the frequency for a Water and Oil system. The volumic fraction of Water is $33\%$. The experiment have been automatised and we proceed as follows for this experiment: starting from rest we record the torque during $60$s and then increase the speed by $0.2$Hz, record one minute and so forth until $4$Hz, which is reached after $20$ minutes. We then decrease the speed until $0$, by steps of $0.2$Hz. 

The torque for the Water and Oil system ($\star$) is at low speed very close to the single-phase case (plotted in the same figure for pure Oil ($\triangle$)). This is consistent with a sum of two layers weakly interacting (small viscosity difference), which basically give the same dimensionnal torque as discussed in figure~\ref{fig:pure}. Increasing the speed, the torque shows up a fast increase around $f=3$Hz (Fig.~\ref{fig:dispersion}a), corresponding to the point at which the liquids start to fully disperse. At the speed of $4$Hz, the torque per unit mass of fluid to stirr is $0.27$N.m.kg$^{-1}$  whereas it is $0.15$N.m.kg$^{-1}$ for pure Oil. This corresponds to an increase of $70\%$.

\begin{figure}
\centering
\includegraphics[width=75mm]{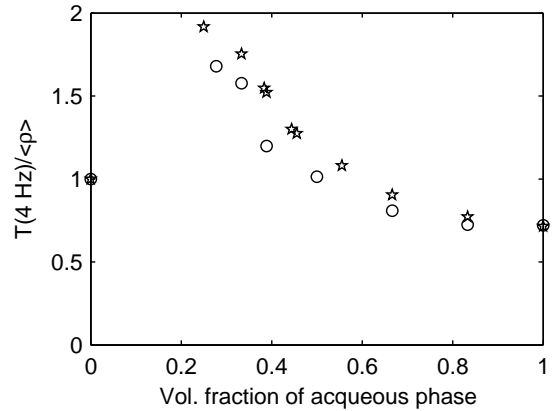}
\caption{Massic torque at $4$Hz, normalised by the pure Oil case {\em vs.} volume fraction of Water ($\circ$) and NaI ($\star$).}
\label{fig:tableau}
\end{figure}

When decreasing the speed, there seems to be hysteresis in the system. The torques on the return path are slightly higher than during the increase. The end of the hysteresis loop seems to be at $f\simeq2$Hz. Indeed, visual observation confirms that the fluids are still dispersed, even below $2$Hz. Similar behaviours are present for other concentrations we studied (see figure~\ref{fig:tableau}), with less hysteresis for the lowest oil concentration. To better understand this hysteresis, we performed the following long-time experiments. Starting from a two-layer system at rest, we increase quickly the rotation up to $f=3.5$Hz and record then the torque (figure~\ref{fig:dispersion}b). The torque gradualy increase during the $600$s of the experiments. The final stationary state is not reached before the end of these $600$s. We plot also this experiment in figure~\ref{fig:dispersion}a as a parametric points cloud with time as parameter. One can notice that the torque at the beginning is the one for Oil, and gradually reaches the upper branch.

The complementary experiment (figure~\ref{fig:dispersion}c) consists of lowering the speed to $2.8$Hz after $20$ minutes. It shows a slow decrease and the final state is not on the lower branch (figure~\ref{fig:dispersion}a): the two-layers system is not the one stable above $2.8$Hz. The time scales involved in the dispersion and separation processes are thus very long, the latter being even longer (figure~\ref{fig:dispersion}b-c). Complicated behaviours including periodic oscillations (with a period of $20\,$min) have been observed in between and requires further investigation.

We report in figure~\ref{fig:tableau} the evolution with the volume fraction of the acqueous phase of the torque per unit mass of fluid to stirr at $4$Hz. The data have been normalised by that of pure Oil ($0.15$N.m.kg$^{-1}$). This quantity is maximum for a volume fraction of the acqueous phase around $33\%$ and reaches a factor $2$. Similar behaviours have been evidenced recently in liquid-liquid turbulent pipe flow experiments (Piela et al., 2006).

\section{Conclusion and perspectives}
We have built a facility with well-controlled parameters to study liquid-liquid dispersions in a turbulent flow. In a first step, the single-phase flow has been characterized in terms of scaling of the dissipation, and also in terms of velocity and Reynolds stress profiles. We measured the turbulent fluctuations intensity who are a key feature for the dispersion mechanisms. Further investigations on the scaling and threshold for first instability for the counterrotating monophasic flow will be performed, on a wider Reynolds number range, using glycerol.

\begin{figure}
\centering
\includegraphics[width=75mm]{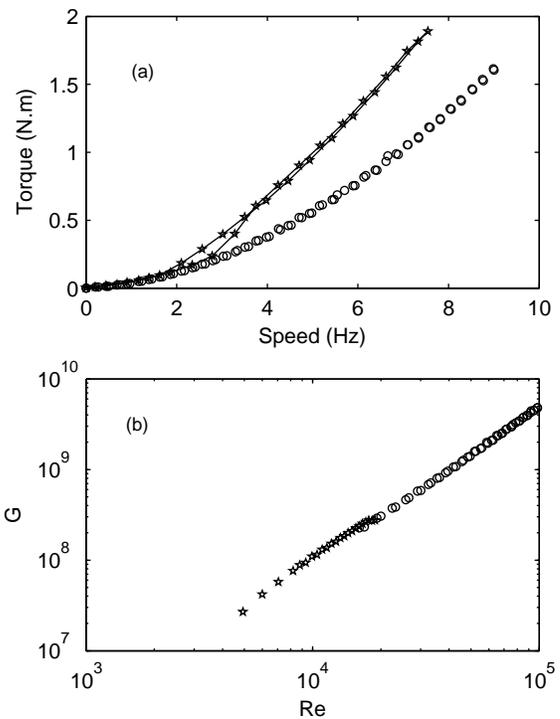}
\caption{(a): Torque \emph{vs.} rotation frequency for pure Water ($\circ$) and $50\%$ Water in Oil ($\star$). (b): Dimensionless Torque $G$ \emph{vs.} $Re$, molecular value for pure Water ($\circ$), and effective values proposed by Pal (2001) for $50\%$ Water in Oil ($\star$).}
\label{fig:zob}
\end{figure}

A global measurement such as in-line torque measurement is able to characterize the state of the system in terms of whether it is dispersed or not. Data presented here for the torques in the dispersions were still limited by old motors with low power. Further work have to be done to explore the scaling of the torque for turbulent dispersions. We plot in figure~\ref{fig:zob} a first result obtained for $50\%$ Water in Oil. The torque in the dispersed state seems to follow a nice scaling at higher velocities. In order to check the idea of a proper "effective molecular viscosity", we try the one proposed by Pal (2001): for viscosity ratio 3 (oil drops in water) and for fraction 0.5, the effective viscosity is $5.9 \times 10^{-6}$m$^2$.s$^{-1}$. It seems then possible to collapse the points onto a single curve when one uses this effective viscosity to compute the Reynolds number and to normalise the torque in the dispersed state (figure~\ref{fig:zob}b). Notice that in fact we do not have enough points at low values of Reynolds numbers for newtonian fluids to really compare this rescaling with consistent data.

In order to understand the interaction between the different scales, we are now working on edge detection algorithm to study the drop size distribution with the LIF images.

\section{References}



Budwig, R., 1994,"Refracive index matching methods for liquid flow investigations", \emph{Exp. in Fluids}, Vol. 17, pp. 350-355.

Dubrulle, B., Dauchot, O., Daviaud, F., Longaretti, P.-Y., Richard, D. and Zahn, J.-P., 2005, "Stability and turbulent transport in Taylor--Couette flow from analysis of experimental data", \emph{Phys. Fluids}, Vol. 17, pp. 095103.

Dubrulle, B., and Hersant, F., 2002, "Momentum transport and torque scaling in Taylor--Couette flow from an analogy with turbulent convection", \emph{Eur. Phys. J. B}, Vol. 26, pp. 379-386.

Esser, A., and Grossmann, S., 1996, "Analytic expression for Taylor–-Couette stability boundary", \emph{Phys. Fluids}, Vol. 8, pp. 1814-1819.

Mari\'e, L., and Daviaud, F., 2004, "Experimental measurement of the scale-by-scale momentum transport budget in a turbulent shear flow", \emph{Phys. Fluids}, Vol. 16, pp. 457-461. 

Narrow, T. L., Yoda, M., and Abdel-Khalik, S. I., 2000, "A simple model for the refractive index of sodium iodide acqueous solutions", \emph{Exp. in Fluids}, Vol. 28, pp. 282-283.

Pal, R., 2001, "Novel viscosity equations for emulsions of two immiscible liquids", \emph{Journal of Rheology}, Vol. 45, pp. 509-520.

Piela, K., Delfos, R., Ooms, G., Westerweel, J., Oliemans, R. V. A., and Mudde, R. F., 2006, "Experimental investigation of phase inversion in an Oil-Water flow through a horizontal pipe loop", \emph{Int. J. Multiphase Flows}, Vol. 32, pp. 1087-1099.

Portela, L. M., and Oliemans, R. V. A., 2006, "Possibilities and Limitations of Computer Simulations of Industrial Turbulent Dispersed Multiphase Flows", \emph{Flow Turbulence Combust.}, Vol. 77, pp. 381-403.

Wang, L., Olsen, M. G., and Vigil, R. D., 2005, "Reappearance of azimuthal waves in turbulent Taylor-Couette flow at large aspect ratio", \emph{Chem. Eng. Sci.}, Vol. 60, pp. 5555-5568.

Westerweel, J., 1993, "Digital particle image velocimetry", Ph.D. Thesis, Delft University of Technology, Delft, Netherlands.


\end{document}